\documentstyle[12pt,psfig]{article}
\begin{document}
\baselineskip=16pt

%%%%%%%%%%%%%%%%%%%%%%%%%%%%%%%%%%%%%%%%%%%%%%%%%%%%%%%%%%%%%%%%%%%
%%%%%%%%%%%%%%%%%%%%%%%%%%%%%%%%%%%%%%%%%%%%%%%%%%%%%%%%%%%%%%%%%%%
\makeatletter
\newcommand{\mat}[4]{\left(\begin{array}{cc}{#1}&{#2}\\{#3}&{#4}\end{array}
\right)}
\newcommand{\matr}[2]{\left(\begin{array}{c}{#1}\\{#2}
\end{array}\right)}           

\def\ra{\rightarrow}
\def\ne{\hbox{$\nu_e$ }}
\def\nm{\hbox{$\nu_\mu$ }}
\def\nt{\hbox{$\nu_\tau$ }}
\def\ns{\hbox{$\nu_{s}$ }}
\def\nx{\hbox{$\nu_x$ }}
\def\Nt{\hbox{$N_\tau$ }}
\def\nr{\hbox{$\nu_R$ }}
\def\O{\hbox{$\cal O$ }}
\def\L{\hbox{$\cal L$ }}
%%%%%%%%%%%%%%%%%%%%%%%%%%%
%%%%%%%%%This defines et al., i.e., e.g., cf., etc.
\def\ie{\hbox{\it i.e., }}        \def\etc{\hbox{\it etc. }}
\def\eg{\hbox{\it e.g., }}        \def\cf{\hbox{\it cf.}}
\def\etal{\hbox{\it et al., }}
%%%%%%%%%%%%%%%%
\def\neus{\hbox{neutrinos }}
\def\gau{\hbox{gauge }}
\def\neu{\hbox{neutrino }}
%%%%%%%%%%%%common physics symbols
\def\c{\mathop{\cos \theta }}
\def\s{\mathop{\sin \theta }}
\def\tr{\mathop{\rm tr}}
\def\Tr{\mathop{\rm Tr}}
\def\Im{\mathop{\rm Im}}
\def\Re{\mathop{\rm Re}}
\def\bR{\mathop{\bf R}}
\def\bC{\mathop{\bf C}}
\def\eq#1{{eq. (\ref{#1})}}
\def\Eq#1{{Eq. (\ref{#1})}}
\def\Eqs#1#2{{Eqs. (\ref{#1}) and (\ref{#2})}}
\def\Eqs#1#2#3{{Eqs. (\ref{#1}), (\ref{#2}) and (\ref{#3})}}
\def\Eqs#1#2#3#4{{Eqs. (\ref{#1}), (\ref{#2}), (\ref{#3}) and (\ref{#4})}}
\def\eqs#1#2{{eqs. (\ref{#1}) and (\ref{#2})}}
\def\eqs#1#2#3{{eqs. (\ref{#1}), (\ref{#2}) and (\ref{#3})}}
\def\eqs#1#2#3#4{{eqs. (\ref{#1}), (\ref{#2}), (\ref{#3}) and (\ref{#4})}}
\def\fig#1{{Fig. (\ref{#1})}}
\def\partder#1#2{{\partial #1\over\partial #2}}
\def\secder#1#2#3{{\partial^2 #1\over\partial #2 \partial #3}}
\def\bra#1{\left\langle #1\right|}
\def\ket#1{\left| #1\right\rangle}
\def\VEV#1{\left\langle #1\right\rangle}
\let\vev\VEV
\def\gdot#1{\rlap{$#1$}/}
\def\abs#1{\left| #1\right|}
\def\pri#1{#1^\prime}
\def\ltap{\raisebox{-.4ex}{\rlap{$\sim$}} \raisebox{.4ex}{$<$}}
\def\gtap{\raisebox{-.4ex}{\rlap{$\sim$}} \raisebox{.4ex}{$>$}}
\def\lsim{\raise0.3ex\hbox{$\;<$\kern-0.75em\raise-1.1ex\hbox{$\sim\;$}}}
\def\gsim{\raise0.3ex\hbox{$\;>$\kern-0.75em\raise-1.1ex\hbox{$\sim\;$}}}
\def\half{{1\over 2}}
%%%%%%%%%%%%%%%%%%%

\def\bea{\begin{eqnarray}}
\def\ea{\end{array}}
\def\beq{\begin{equation}}
\def\eeq{\end{equation}}
%%%%%%%%%%%%%%%%%%%%
\def\bef{\begin{figure}}
\def\eef{\end{figure}}
\def\bet{\begin{table}}
\def\eeqt{\end{table}}
\def\bea{\begin{eqnarray}}
\def\ba{\begin{array}}
\def\ea{\end{array}}
\def\bi{\begin{itemize}}
\def\ei{\end{itemize}}
\def\ben{\begin{enumerate}}
\def\eeqn{\end{enumerate}}
\def\ra{\rightarrow}
\def\ot{\otimes}
\def\ap#1#2#3{           {\it Ann. Phys. (NY) }{\bf #1} (19#2) #3}
\def\arnps#1#2#3{        {\it Ann. Rev. Nucl. Part. Sci. }{\bf #1} (19#2) #3}
\def\cnpp#1#2#3{        {\it Comm. Nucl. Part. Phys. }{\bf #1} (19#2) #3}
\def\apj#1#2#3{          {\it Astrophys. J. }{\bf #1} (19#2) #3}
\def\app#1#2#3{          {\it Astropart. Phys. }{\bf #1} (19#2) #3}
\def\asr#1#2#3{          {\it Astrophys. Space Rev. }{\bf #1} (19#2) #3}
\def\ass#1#2#3{          {\it Astrophys. Space Sci. }{\bf #1} (19#2) #3}
\def\aa#1#2#3{          {\it Astron. \& Astrophys.  }{\bf #1} (19#2) #3}
\def\apjl#1#2#3{         {\it Astrophys. J. Lett. }{\bf #1} (19#2) #3}
\def\ap#1#2#3{         {\it Astropart. Phys. }{\bf #1} (19#2) #3}
\def\ass#1#2#3{          {\it Astrophys. Space Sci. }{\bf #1} (19#2) #3}
\def\jel#1#2#3{         {\it Journal Europhys. Lett. }{\bf #1} (19#2) #3}
\def\ib#1#2#3{           {\it ibid. }{\bf #1} (19#2) #3}
\def\nat#1#2#3{          {\it Nature }{\bf #1} (19#2) #3}
\def\nps#1#2#3{        {\it Nucl. Phys. B (Proc. Suppl.) }{\bf #1} (19#2) #3} 
\def\np#1#2#3{           {\it Nucl. Phys. }{\bf #1} (19#2) #3}
\def\pl#1#2#3{           {\it Phys. Lett. }{\bf #1} (19#2) #3}
\def\pr#1#2#3{           {\it Phys. Rev. }{\bf #1} (19#2) #3}
\def\prep#1#2#3{         {\it Phys. Rep. }{\bf #1} (19#2) #3}
\def\prl#1#2#3{          {\it Phys. Rev. Lett. }{\bf #1} (19#2) #3}
\def\pw#1#2#3{          {\it Particle World }{\bf #1} (19#2) #3}
\def\ptp#1#2#3{          {\it Prog. Theor. Phys. }{\bf #1} (19#2) #3}
\def\jppnp#1#2#3{         {\it J. Prog. Part. Nucl. Phys. }{\bf #1} (19#2) #3}
\def\rpp#1#2#3{         {\it Rep. on Prog. in Phys. }{\bf #1} (19#2) #3}
\def\ptps#1#2#3{         {\it Prog. Theor. Phys. Suppl. }{\bf #1} (19#2) #3}
\def\rmp#1#2#3{          {\it Rev. Mod. Phys. }{\bf #1} (19#2) #3}
\def\zp#1#2#3{           {\it Zeit. fur Physik }{\bf #1} (19#2) #3}
\def\fp#1#2#3{           {\it Fortschr. Phys. }{\bf #1} (19#2) #3}
\def\Zp#1#2#3{           {\it Z. Physik }{\bf #1} (19#2) #3}
\def\Sci#1#2#3{          {\it Science }{\bf #1} (19#2) #3}
\def\n.c.#1#2#3{         {\it Nuovo Cim. }{\bf #1} (19#2) #3}
\def\r.n.c.#1#2#3{       {\it Riv. del Nuovo Cim. }{\bf #1} (19#2) #3}
\def\sjnp#1#2#3{         {\it Sov. J. Nucl. Phys. }{\bf #1} (19#2) #3}
\def\yf#1#2#3{           {\it Yad. Fiz. }{\bf #1} (19#2) #3}
\def\zetf#1#2#3{         {\it Z. Eksp. Teor. Fiz. }{\bf #1} (19#2) #3}
\def\zetfpr#1#2#3{    {\it Z. Eksp. Teor. Fiz. Pisma. Red. }{\bf #1} (19#2) #3}
\def\jetp#1#2#3{         {\it JETP }{\bf #1} (19#2) #3}
\def\mpl#1#2#3{          {\it Mod. Phys. Lett. }{\bf #1} (19#2) #3}
\def\ufn#1#2#3{          {\it Usp. Fiz. Naut. }{\bf #1} (19#2) #3}
\def\sp#1#2#3{           {\it Sov. Phys.-Usp.}{\bf #1} (19#2) #3}
\def\ppnp#1#2#3{           {\it Prog. Part. Nucl. Phys. }{\bf #1} (19#2) #3}
\def\cnpp#1#2#3{           {\it Comm. Nucl. Part. Phys. }{\bf #1} (19#2) #3}
\def\ijmp#1#2#3{           {\it Int. J. Mod. Phys. }{\bf #1} (19#2) #3}
\def\ic#1#2#3{           {\it Investigaci\'on y Ciencia }{\bf #1} (19#2) #3}
\def\tp{these proceedings}
\def\pc{private communication}
\def\opc{\hbox{{\sl op. cit.} }}
\def\ip{in preparation}
%%%%%%%%%%%%%%%%%%%%%%%%%%%%%%%%%%%%%%%%%%%%%%%%%%%%%%
\hoffset=0.1in
\voffset=-0.3in
\renewcommand{\baselinestretch}{2}
\renewcommand{\baselinestretch}{1}
%\newcommand{\resection}[1]{\setcounter{equation}{0}\section{#1}}
%\newcommand{\appsection}{\addtocounter{section}{1} \setcounter{equation}{0}
%                         \section*{Appendix \Alph{section}}}
%\renewcommand{\theequation}{\arabic{equation}}
%\textwidth 200mm
%\hsize=17cm
%\textheight 230mm
% from here remove % to produce the right format
\topmargin -0.6cm
\oddsidemargin=-0.75cm
\hsize=16.8cm
\setlength{\textheight}{247mm}
%   until here
%\parindent=0.7truecm
%\topmargin -4pt
%\oddsidemargin=-0.4truecm
%\evensidemargin=-0.4truecm

%%%%%%%%%%%%%

\renewcommand{\thefootnote}{\fnsymbol{footnote}}
\def\e{\mbox{e}}
\def\sgn{{\rm sgn}}
\def\gsim{\;
\raise0.3ex\hbox{$>$\kern-0.75em\raise-1.1ex\hbox{$\sim$}}\;}
\def\lsim{\;
\raise0.3ex\hbox{$<$\kern-0.75em\raise-1.1ex\hbox{$\sim$}}\;}
\def\MeV{\rm MeV}
\def\eV{\rm eV}
%\tableofcontents
\thispagestyle{empty}
\begin{titlepage}
%%\today
\begin{center}
\rightline{hep-ph/9605427}
\hfill FTUV/96-32\\
\hfill IFIC/96-37\\
\vskip 0.3cm
\large
{\bf The stability of  the  MSW solution to the solar neutrino problem 
with respect to random matter density perturbations\footnote{
Invited talk presented at ``XXXIst Les Rencontres de Moriond-
Electroweak Interactions and Unified Theories'', Les Arcs, France - 
March 16-23, 1996.\\
This contribution is based on the paper [1] done in collaboration with
H. Nunokawa, V. Semikoz and J. W. F. Valle.
}}
\end{center}
\normalsize

\begin{center}
{\bf Anna Rossi}
\footnote{
E-mail: rossi@evalvx.ific.uv.es, rossi@ferrara.infn.it},
\end{center}
\begin{center}
{\it Instituto de F\'{\i}sica Corpuscular - C.S.I.C.\\
Departament de F\'{\i}sica Te\`orica, Universitat de Val\`encia\\}
{\it 46100 Burjassot, Val\`encia, SPAIN         }\\
\vglue 2.cm
\end{center}

\begin{center}
{\bf Abstract}
\end{center}

We present a generalization of the resonant 
neutrino conversion in matter, 
including a random component in the matter density profile. 
The  study is focused on the effect of such matter 
perturbations upon both large and 
small mixing angle MSW solutions to the solar neutrino problem. 
This is carried out both for the active-active $\nu_e \ra \nu_{\mu,\tau}$ 
as well as active-sterile $\nu_e \ra \nu_s$ conversion channels. 
We find that the small mixing MSW solution is much more stable 
(especially in $\delta m^2$) than the large mixing solution. 
Future solar neutrino 
experiments, such as  Borexino, could probe solar matter density noise 
at the few percent level.

\vfill

\end{titlepage}
\renewcommand{\thefootnote}{\arabic{footnote}}
\setcounter{footnote}{0}
\newpage

{\bf 1.} The comparison among the 
present experimental results on the observation of the solar neutrinos 
strongly points to a deficit of  neutrino flux (dubbed 
the Solar Neutrino 
Problem (SNP)).  
The most recent averaged data of 
the chlorine \cite{cl}, gallium \cite{ga,sa} and Kamiokande 
\cite{k} experiments are:
\beq
\label{data}
R_{Cl}^{exp}= (2.55 \pm 0.25) \mbox{SNU}, \,\,\,\,
R_{Ga}^{exp}= (74 \pm 8) \mbox{SNU 
\footnote{For the gallium result we have taken the weighted average of GALLEX 
$R^{exp}_{Ga}= (77\pm8\pm5)$SNU\cite{ga} and SAGE 
$R^{exp}_{Ga}= (69\pm 11\pm 6)$SNU \cite{sa} measurements.}
}, \,\,\,\
R_{Ka}^{exp}= (0.44 \pm 0.06) R_{Ka}^{BP95} 
\eeq 
where  $R_{Ka}^{BP95}$ is the prediction according to the   
most recent Standard Solar Model (SSM)by 
Bahcall-Pinsonneault (BP95)\cite{SSM}  . 

In particular the SNP is now understood as the strong 
deficit of the beryllium neutrinos \cite{CF}. On the other hand, 
the high energy boron neutrinos are moderately suppressed, while 
the low energy ones are almost undepleted.  All this seems to imply 
that any astrophysical solution fails \cite{CF,BFL} in reconciling 
the experimental data. 

From the particle physics point of view, however, 
the resonant neutrino conversion due to the neutrino 
interactions with constituents of the solar material  
(the Mikheyev-Smirnov-Wolfenstein  (MSW) effect) \cite{MSW} offers the best 
explanation of the present experimental situation. 
This scenario provides an extremely good data fit in the small mixing 
region with  $\delta m^2 \simeq 10^{-5}$eV$^2$ and 
$\sin^2 2 \theta \simeq 10^{-3} \div 10^{-2}$ \cite{FIT,smirnov,Cala}. 
Moreover, the study of the MSW effect has revealed its stability 
against possible 
changes of the SSM input parameters \cite{smirnov} 
especially in the $\delta m^2$ parameter. 

 This  talk deals with the stability of the MSW solution with 
respect to the possible presence of random perturbations in the solar 
matter density \cite{NRSV}. 

We remind that 
in Ref.\cite{KS} the effect of periodic matter density perturbations 
added to a mean  matter density $\rho$
upon resonant neutrino conversion was investigated. The major effects 
show up when the fixed frequency  of the perturbation is close 
to the neutrino oscillation eigen-frequency, and for rather large amplitude 
values ($\sim 10\div 20\%$), giving rise to the parametric effects \cite{KS}.  
There are also a number of papers which address similar effects by different 
approaches \cite{AbadaPetcov,BalantekinLoreti}. 

Here we consider the effect of random 
matter density perturbations $\delta \rho(r)$, characterised by an 
{\sl arbitrary} wave number $k$, 
\beq
\delta \rho (r) = \int dk \delta \rho(k)\sin kr  \:,
\eeq
rather than a periodic or regular perturbation. 
The effect of solar density as well as solar magnetic field 
fluctuations upon neutrino spin-flavour conversions has  
also been considered in Ref. \cite{BalantekinLoreti},
using somewhat different methods.

Moreover, as in  Ref.\cite{BalantekinLoreti}, 
we assume that the perturbation $\delta \rho$ 
has Gaussian distribution.
For small inhomogeneities, the spatial correlation function 
$\langle \xi^2 \rangle$ can be taken as 
\beq\label{correlator}
\langle \delta \rho(r_1)\delta \rho(r_2)\rangle = 2\rho^2\langle 
\xi^2\rangle L_0 \delta (r_1 - r_2)\, , \,\,\,\,\,\,
\langle 
\xi^2\rangle \equiv 
\frac{\langle \delta \rho^2\rangle}{\rho^2}\, , 
\eeq
whose correlation length $L_0$ obeys the following relation:
\beq\label{size}
l_{{free}} \ll L_0 \ll \lambda_m
\eeq
where   
$l_{\rm free}= (\sigma n_0)^{-1}$ is the mean free path 
of the electrons 
in the solar medium\footnote{
For Coulomb interactions, 
the cross-section $\sigma$  is determined by the classical radius of 
electron $r_{0e} =e^2/m_ec^2\sim 2 \times 10^{-13}$cm, resulting in 
$l_{\rm free}\sim 10\,\mbox{cm}$ for a solar mean density 
$n_0 \sim 10^{24}$cm$^{-3}$.}, and $\lambda_m$ is the neutrino 
matter wave length. 
 The  lower bound, is dictated by the 
hydro-dynamical approximation used later, whereas 
the upper bound expresses the fact that the scale of 
fluctuations should be much smaller than $\lambda_m$ (the characteristic 
neutrino propagation length scale),  as indeed the 
Eq. (\ref{correlator}) requires.
For the sake of discussion, in the following 
%n order tto ensure that the
%correlation length $L_0$ is smaller than the neutrino 
%wave length in the Sun (cfr. Eq. (\ref{size})), 
we choose to adjust $L_0$ as follows:
\beq\label{L0}
L_0 = 0.1 \times  \lambda_m .
\eeq
The SSM in itself cannot account for the existence of 
density perturbations, since it is based on hydrostatic 
evolution equations. On the other hand, the present 
helioseismology observations cannot 
exclude the existence of few percent 
level of matter density fluctuations \cite{dal,gmode}. 
Therefore, in what follow we assume, on phenomenological grounds, 
such  levels for $\xi$, up to 8\%. 

Before generalizing the MSW scenario,
 accounting for the presence in the interior of the 
sun of  such matter density fluctuations, first
we  give a quick reminder to the main features of the MSW effect. 

\vspace{0.5cm}

{\bf 2.} 
The resonant conversion of neutrinos in a matter background is due
to the coherent neutrino scattering off matter constituents \cite{MSW}. 
This determines an effective matter potential $V$ for neutrinos.
In the rest frame of the unpolarised matter, the potential 
is given, in the framework of the Standard Model, by 
\beq\label{poten}
V = \frac{\sqrt{2}G_F}{m_p} \rho Y
\eeq
where $G_F$ is the Fermi constant and 
$Y$ is a number  which depends on the \neu type and on the chemical 
content of the medium. More precisely, $Y= Y_e - \frac{1}{2}Y_n$ for 
the $\nu_e$ state, $Y= -\frac{1}{2}Y_n$ for \nm and \nt and $Y=0$ 
for the sterile $\nu_s$ state, where $Y_{e,n}$ denotes the electron and 
neutron number per nucleon. 
Let us note the dependence of $V$  on the matter density $\rho$
 for which one  usually consider the {\it smooth} distribution, 
as given by the SSM \cite{SSM,turck,CDF}.

Once assumed that there exists a non-vanishing mass difference 
$\delta m^2$ between two different neutrino states and a 
non-vanishing neutrino mixing $\theta$ in vacuum, 
the neutrinos $\nu_e$'s,  
created in the inner region of the sun, where the 
$\rho$ distribution is maximal, 
can be completely converted into $\nu_y$ ($y= \mu$, $\tau$ or $s$), 
while travelling to the solar surface. \\
This requires two conditions \cite{MSW}:

1) - the resonance condition. Neutrinos of given energy 
$E$ experience the resonance if the energy splitting in the vacuum 
$\delta m^2 \cos 2 \theta / 2E$ 
is compensated by the effective matter potential 
difference $\Delta V_{ey} = V_e - V_y$. It is helpful to define the 
following dynamical factor $A_{ey}$
\beq
\label{afactor}
A_{ey}(r) = \frac{1}{2} [\Delta V_{ey} (r) 
 - \frac{\delta m^2}{2E} \cos2 \theta]
\eeq
which vanishes at the resonance, $A_{ey}=0$. 
This condition determines 
the value \\  
$\rho_{res} = (m_p \cos 2 \theta /2 \sqrt{2} G_F) (Y_e-Y_y) 
\delta m^2 /E$ which, in turn,  implies a resonance layer $\Delta r$. 

2) - The adiabatic condition. 
At the resonance layer, the neutrino conversion $\nu_e\to \nu_y$ 
is efficient if the propagation is adiabatic.
This can  be nicely expressed 
requiring that the neutrino wavelenght $\lambda_m$ 
be smaller than $\Delta r$ \cite{MSW},
\begin{eqnarray}
\label{alfamsw}
\alpha_r &  = & \Delta r/(\lambda_m)_{res} \equiv 
\frac{\delta m^2 \sin^2 2 \theta R_0}{4\pi E \cos 2\theta} > 1\, ,
\,\,\,
R_0 \approx 0.1 R_{\odot} \, ,\\
\lambda_m & = & \frac{\pi}{\sqrt{ A_{ey}^2 +  
(\delta m^2)^2\sin^2 2\theta/16 E^2}}\,, \,\,\,\,\,\, \Delta r = 2 \rho_{res} 
\tan 2\theta 
|\mbox{d}\rho/\mbox{d}r|^{-1}\,.  \nonumber 
\end{eqnarray}

\vspace{0.5cm}

{\bf 2.}
Now we re-formulate the evolution eqaution 
for the neutrino 
accounting for a fluctuation 
term $\delta \rho$ superimposed to the main profile $\rho$. 
The perturbation level $\xi =\frac{\delta \rho}{\rho}$ 
induces a corresponding 
random component for the matter potential of the form 
$\Delta V_{ey} \xi$.
The evolution for the $\nu_e-\nu_y$ 
system is governed by 
\beq
\label{ev1}
i \frac{d}{dt}\matr{\nu_e} {\nu_y} =
\mat{H_{e}}  {H_{e y}} 
             {H_{ey}}  {H_{y}}\matr{\nu_e} {\nu_y}, 
\eeq
where the entries of the Hamiltonian matrix are given by
%\footnote{In the Hamiltonian matrix, a term proportional 
%to the identity has been removed.}
 \begin{eqnarray}
\label{matdef}
 & & H_e=  2 [A_{ey}(t) + \tilde{A}_{ey}(t)], ~~~~ H_y=0, 
~~~~ H_{ey}=\frac{\delta m^2}{4E} \sin2 \theta, 
\nonumber\\ 
  & & A_{ey}(t)  =  \frac{1}{2} [\Delta V_{ey}(t) 
 - \frac{\delta m^2}{2E} \cos2 \theta], ~~~~~
\tilde{A}_{ey}(t) = \frac{1}{2} \Delta V_{ey}(t) \xi
\end{eqnarray}
Here the matter potential for the active-active 
$\nu_e\ra \nu_{\mu,\tau}$ conversion 
reads 
\beq
\label{vex}
\Delta V_{e\mu (\tau)}(t) = \frac{\sqrt{2} G_F}{m_p} \rho(t) (1-Y_n)
\eeq
or alternatively in case $\nu_e\ra\nu_{s}$  
\beq
\label{vexs}
\Delta V_{es}(t) = \frac{\sqrt{2} G_F}{m_p} \rho(t) (1-\frac{3}{2}Y_n)
\eeq
(the neutral matter relation $Y_e =1-Y_n$ has been used). 

The above system  can be rewritten in terms of the following 
equations:
\begin{eqnarray}
\label{sys}
\dot{P}(t)& =& 2H_{e{y}} I(t) \nonumber \\
\dot{R}(t)& = & - H_e(t)I(t) \nonumber \\
\dot{I}(t)& = & H_e(t)R(t) - H_{e{y}} (2P(t) - 1) 
\end{eqnarray} 
where $P = | \nu_e |^2$ is the $\nu_e$ survival probability,  
$R \equiv$ Re$(\nu_y \nu_e^*) $ and 
$I \equiv$ Im$(\nu_y \nu_e^*) $ with 
the corresponding initial conditions 
$P(t_0) = 1, \,I(t_0)=0, \,R(t_0) = 0$.
The Eqs. (\ref{sys}) have to be averaged (see \cite{NRSV} for more details) 
over the 
random density distribution, taking into account that 
for the random component we  have:
\begin{eqnarray}
\label{matcorrel}
& & \langle \tilde{A}_{ey}^{2n+1} \rangle  = 0, ~~~~~~
\langle \tilde{A}_{ey}(t)\tilde{A}_{ey}(t_{1}) \rangle = 
\kappa\delta (t - t_{1}) ,\\
\label{den_noise}
& & \kappa(t)=  \langle \tilde{A}_{ey}^2(t)\rangle L_0 = \frac{1}{2} 
\Delta V^2_{ey}(t)
\langle \xi^2\rangle L_0 .
\end{eqnarray}
The noise-averaged version of the system (\ref{sys}) reads as :
\begin{eqnarray}
\label{sys1}
\dot{\cal{P}}(t) &= &2 H_{ey} \cal{I}(t) \nonumber \\
\dot{\cal{R}}(t) & = & -2A_{ey}(t) \cal{I}(t) -2 \kappa(t)\cal{R}(t) 
 \nonumber \\
\dot{\cal{I}}(t) & = & 2A_{ey}(t) \cal{R}(t) -2 \kappa(t)\cal{I}(t) 
- H_{ey} (2 \cal{P}(t)-1) .
\end{eqnarray}
where clearly $\langle P(t)\rangle = \cal {P}(t)$, 
$\langle R(t)\rangle = \cal {R}(t)$, $\langle I(t)\rangle = \cal {I}(t)$. 
As expected the system of equations (\ref{sys1}) explicitly 
exhibits the noise parameter $\kappa$. 

It is now possible to envisage the main effects due to the presence of the 
random field $\delta \rho$ upon the MSW scenario.
Now the `` dynamics '' is governed by one more quantity i.e. the 
noise parameter $\kappa$, besides the factor $A_{ey}$. Actually, the 
quantity $\kappa$ can be given the meaning of energy quantum associated 
with the matter density perturbation. 
However, let us note that the MSW resonance condition,  
i.e. $A_{ey}(t) =0 $ 
remains unchanged, 
due to the random nature of the matter perturbations. 
The comparison between  the noise parameter $\kappa$ in 
\Eq{den_noise} and $A_{ey}(t)$ shows that 
$\kappa(t) < A_{ey}(t)$, for $\xi \lsim$
few \%, except at the resonance region. 
As a result, the density 
perturbation can have its maximal effect just at the resonance. 
Furthermore, one can find the analogous of condition 2) 
(see Eq. (\ref{alfamsw}) for the noise to give rise to sizeable 
effects. Since the noise term gives rise to a damping term in the system 
(\ref{sys1}), it follows that the corresponding noise length scale 
$1/\kappa$ be much smaller than the thickness of the resonance 
layer $\Delta r$. In other words, the 
following {\it adiabaticity} condition 
\beq
\label{adiab}
\tilde{\alpha}_r= \Delta r\, \kappa_{res} > 1 ,
\eeq
is also necessary. 
There is a simple relation between the 
two adiabaticity parameters $\alpha_r$ (cfr. (\ref{alfamsw})) 
and $\tilde{\alpha}_r$: 
\beq
\label{alfa}
\tilde{\alpha}_r \approx \alpha_r\, \frac{\xi^2}{\tan^2 2\theta} \, .
\eeq
For the range of parameters we are considering, $\xi \sim 10^{-2}$ 
and $\tan^2 2\theta\geq 10^{-3}-10^{-2}$, and due to the r.h.s of 
(\ref{size}), there results $\tilde{\alpha}_r \leq \alpha_r$. 
This relation  can be 
rewritten as $\kappa_{res} < \delta H_{res}$, where $\delta H_{res}$ 
is the level splitting between the energies of the neutrino mass 
eigenstates at resonance. This shows that the noise energy quantum 
is unable to ``excite'' the system, causing the 
level crossing (even at the resonance) \cite{KS}. 
In other words, it never
violates the MSW adiabaticity condition.  
From Eq. (\ref{alfa}) it follows also that, in the adiabatic regime 
$\alpha_r >1$,  the smaller
the mixing angle value the larger 
the effect of the noise.  Finally, as already noted above,
the MSW non-adiabaticity $\alpha_r <1$ 
is always transmitted to $\tilde{\alpha}_r < 1$. As a result,
under our assumptions the fluctuations are expected to be 
ineffective in the non-adiabatic MSW regime. 

\vspace{0.5cm}

{\bf 3.}
All this preliminary discussion is illustrated in the Fig. 1. 
For definiteness we  take BP95 SSM \cite{SSM}
as  reference model.
We plot  $\cal{P}$ as a function 
of $E/\delta m^2$ for different values of the noise parameter $\xi$.
For comparison, the standard MSW case $\xi=0$ is also shown 
(lower solid curve). 
One can see that in both cases of small and large mixing 
(Fig. 1a and Fig. 1b, respectively), the effect of the matter
density noise is to raise the bottom of the pit (see 
dotted and dashed curves). For example, the enhancement of the survival 
probability can easily reach 20\% for $\xi$ values as small as $4\%$. 
In other words, the noise weakens 
the MSW suppression in the adiabatic-resonant 
regime, whereas its effect is  negligible  in 
the non-adiabatic region, in complete agreement with the results 
of Ref.\cite{BalantekinLoreti}. The relative increase 
of the survival probability $\cal{P}$ is larger for the case 
of small mixing (Fig. 1a) as already guessed on the basis of 
Eq. (\ref{alfa}). 
We have also drawn pictorially  (solid vertical line) the 
position, in the $\cal{P}$ profile, 
where $^7Be$  neutrinos fall in for the relevant 
$\delta m^2 \sim 10^{-5}$ eV$^2$, to visualize 
that these intermediate energy neutrinos are the ones most likely
to be affected by the matter noise. 

\vspace{0.5cm}

{\bf 4.} Let us 
analyse the possible impact of this 
scenario in the determination of solar neutrino parameters
from the experimental data. 
For that we have performed the standard $\chi^2$ fit in the $(\sin^2 
2 \theta, \delta m^2)$ parameter space.
The results of the fitting 
are shown in Fig. 2 where the 90\% 
confidence level (C.L.) areas are drawn  for different 
values of $\xi$. 
Fig. 2a and Fig. 2b refer to the cases of $\nu_e 
\to \nu_{\mu,\tau}$ and  $\nu_e 
\to \nu_{s}$ conversion, respectively. 
One can observe that the small-mixing  region is almost stable, 
with a slight shift 
down of $\delta m^2$ values and a slight shift of  
$\sin ^2 2\theta$ towards larger values. 
The large mixing area is also pretty stable, exhibiting 
the tendency to shift to smaller $\delta m^2$ and $\sin^2 2 \theta$.
The  smaller  $\delta m^2$ values compensate for the 
weakening of the MSW suppression due to the presence of 
matter noise,   so that a larger portion of 
the neutrino energy spectrum can be converted. 
The   $\xi=8\%$ case, considered for the sake of demonstration, 
clearly shows that the small mixing region is 
much more stable than the 
large mixing one even for such large value of the noise. 
Moreover  the strong selective 
$^7$Be neutrino suppression, which is the nice feature of the MSW effect, 
is somewhat degraded by the presence of matter noise.  
Consequently the longstanding conflict between chlorine and Kamiokande data 
is exacerbated and the data fit gets worse. 
Indeed, the presence of the matter 
density noise  makes the data fit a little poorer: 
$\chi^2_{min}= 0.1$  for  $\xi=0$, it 
becomes $\chi^2_{min}= 0.8$ for $\xi=$ 4\% and even 
$\chi^2_{min}= 2$ for $\xi=$8\% for the $\nu_e 
\to \nu_{\mu,\tau}$ transition. 

The same tendency is met in the 
case of transition into a sterile state (Fig. 2b): 
$\chi^2_{min}= 1$  for  $\xi=0$, it 
becomes $\chi^2_{min}= 3.6$ for $\xi=$ 4\% and 
$\chi^2_{min}= 9$ for $\xi=$8\%. 

In conclusion 
we have shown that the MSW 
solution to the SNP exists for any realistic levels of matter density noise 
($\xi\leq 4\%$).  
Moreover the MSW solution is essentially stable in mass ($4\cdot 10^{-6}
\mbox{eV}^2 <\delta m^2< 10^{-5}\mbox{eV}^2$ at 90\% CL), whereas 
the mixing appears more sensitive to the level of fluctuations.

\vspace{0.5cm}

{\bf 5.} 
We can reverse our point of view, wondering whether the solar 
neutrino experiments can be a tool to get information on the 
the level of matter noise in the sun.  
In particular, the 
future Borexino experiment \cite{borex},  
aiming to detect the $^7$Be neutrino flux   could be 
sensitive to the presence of solar matter fluctuations. 
%as the $^7$Be neutrinos are those 
%mostly affected by the presence of matter noise.
In the relevant  MSW parameter region for the noiseless case,  
the Borexino signal cannot be definitely predicted 
(see  Fig. 3a). Within the present allowed C.L. regions (dotted line)   
the expected rate,  $Z_{Be}\!=\!R^{pred}_{Be}/R^{BP95}_{Be}$ (solid lines), 
is in the range $0.2\div 0.7$. 

On the other hand, when the  matter density noise is switched on, e.g. 
 $\xi= 4\%$ (see Fig. 3b), the minimal 
allowed value for $Z_{Be}$ becomes higher, $Z_{Be}\!\geq \!0.4$. 
Hence,  if the MSW mechanism is responsable for the 
solar neutrino deficit and Borexino
experiment  detects a low signal, say $Z_{Be}\lsim 0.3$
(with good accuracy)  this will imply that a 4\% level of matter 
fluctuations in the central region of the sun is unlikely . 
The same argument can be applied to 
\ne $\ra$ \ns resonant conversion, whenever  future 
large detectors such 
as Super-Kamiokande 
and/or the Sudbury Neutrino Observatory (SNO) 
establish through, e.g. the measurement of the charged to neutral 
current ratio,
that  the deficit of solar neutrinos is due to this kind of transition. 
The expected signal in Borexino is very small $Z_{Be} \approx 0.02$ for 
$\xi =0$ (see Fig. 3c). 
On the other hand with $\xi=4\%$, 
the minimum expected Borexino signal is 10 times higher than in the
noiseless case, so that if Borexino detects a rate $Z_{Be} \lsim 0.1$ 
(see Fig. 3d) this would again exclude noise levels above  $4\%$.

Let us notice that Super-Kamiokande and SNO experiments, being sensitive 
only to the higher energy Boron neutrinos, probably 
do not offer similar possibility 
to probe such matter fluctuations in the sun. 

The previous discussion, which certainly deserves a more accurate 
analysis
 involving also the theoretical uncertainties in the 
 $^7$Be neutrino flux, shows the close link between neutrino physics and 
solar physics.

\vspace{0.5cm}

It is a pleasure to  thank   N. Yahlali and H. Nunokawa for reading 
the manuscript. 
This work has been supported by 
the grant N. ERBCHBI CT-941592 of the Human Capital and Mobility 
Program.

\newpage
%%%%%%%%%%%%%%%%%%% Fig. 1a and 1b %%%%%%%%%%%%%%%%%%%%%%%%%%%%%%%
\hglue -1.5cm
\psfig{file=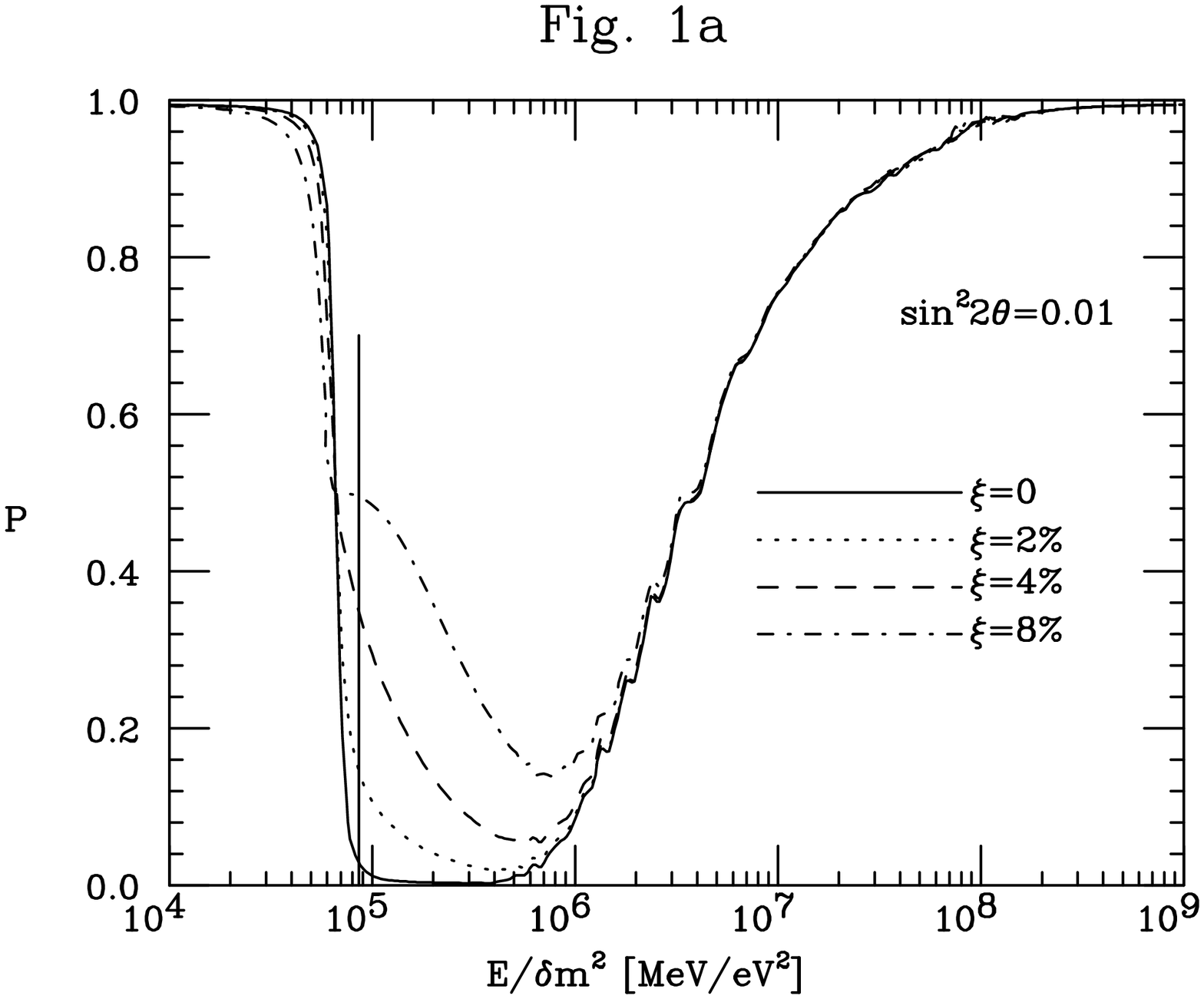,height=8.5cm,width=9.5cm,angle=90}
{\vglue -8.5cm 
\hglue 7cm
\psfig{file=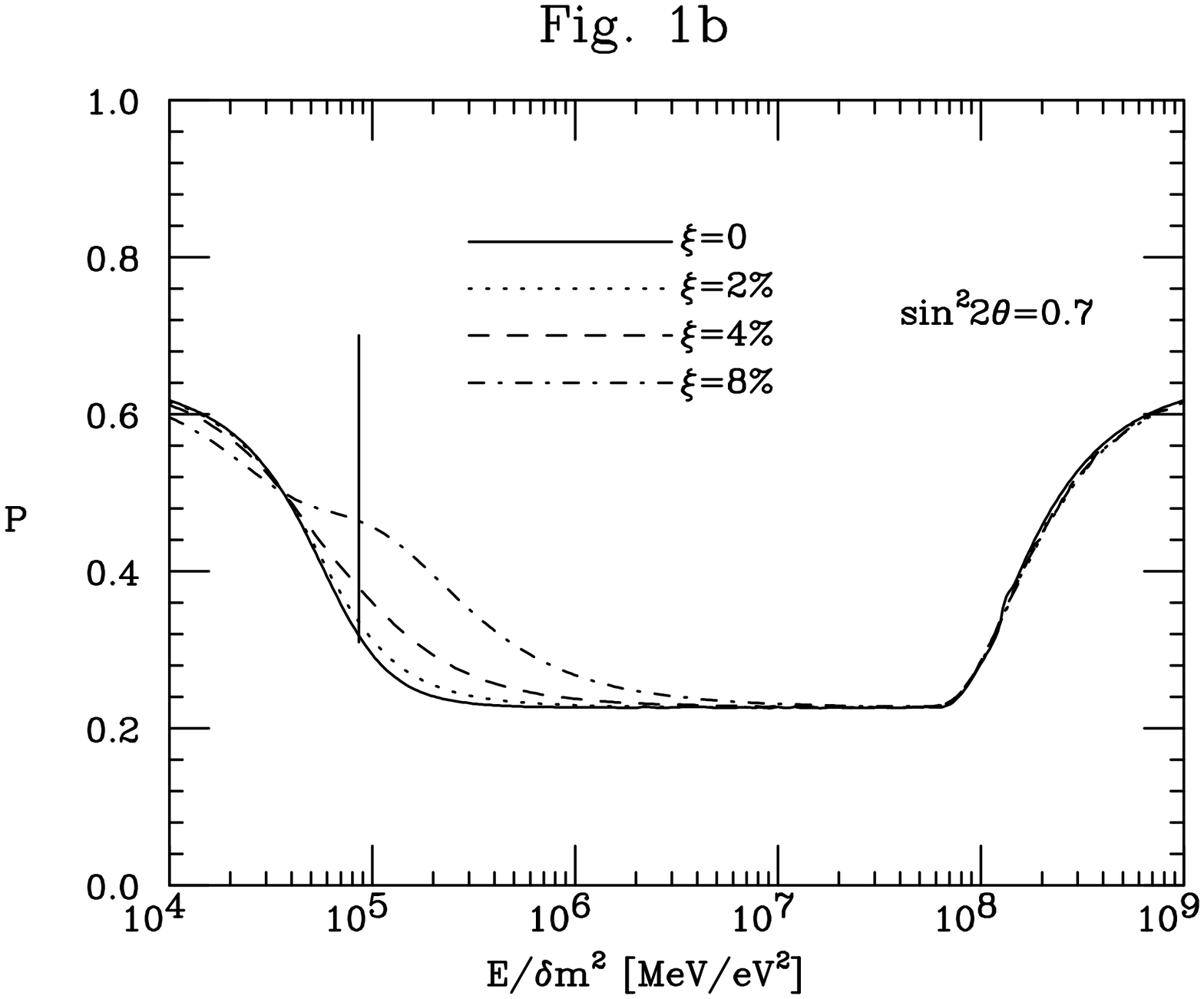,height=8.5cm,width=9.5cm,angle=90} 
}
\noindent
Fig. 1: The averaged solar neutrino survival probability {\cal P} 
versus $E/\delta m^2$ for small mixing angle, $\sin^2 2\theta=0.01$, (Fig. 1a) 
and for large mixing angle, $\sin^2 2\theta=0.7$, (Fig. 1b). 
The different curves refer to different values of matter noise 
level $\xi$ as indicated.
%%%%%%%%%%%%%%%%%%% Fig. 2a and 2b %%%%%%%%%%%%%%%%%%%%%%%%%%%%%%%
\vglue 2.0cm
\hglue -1.5cm
\psfig{file=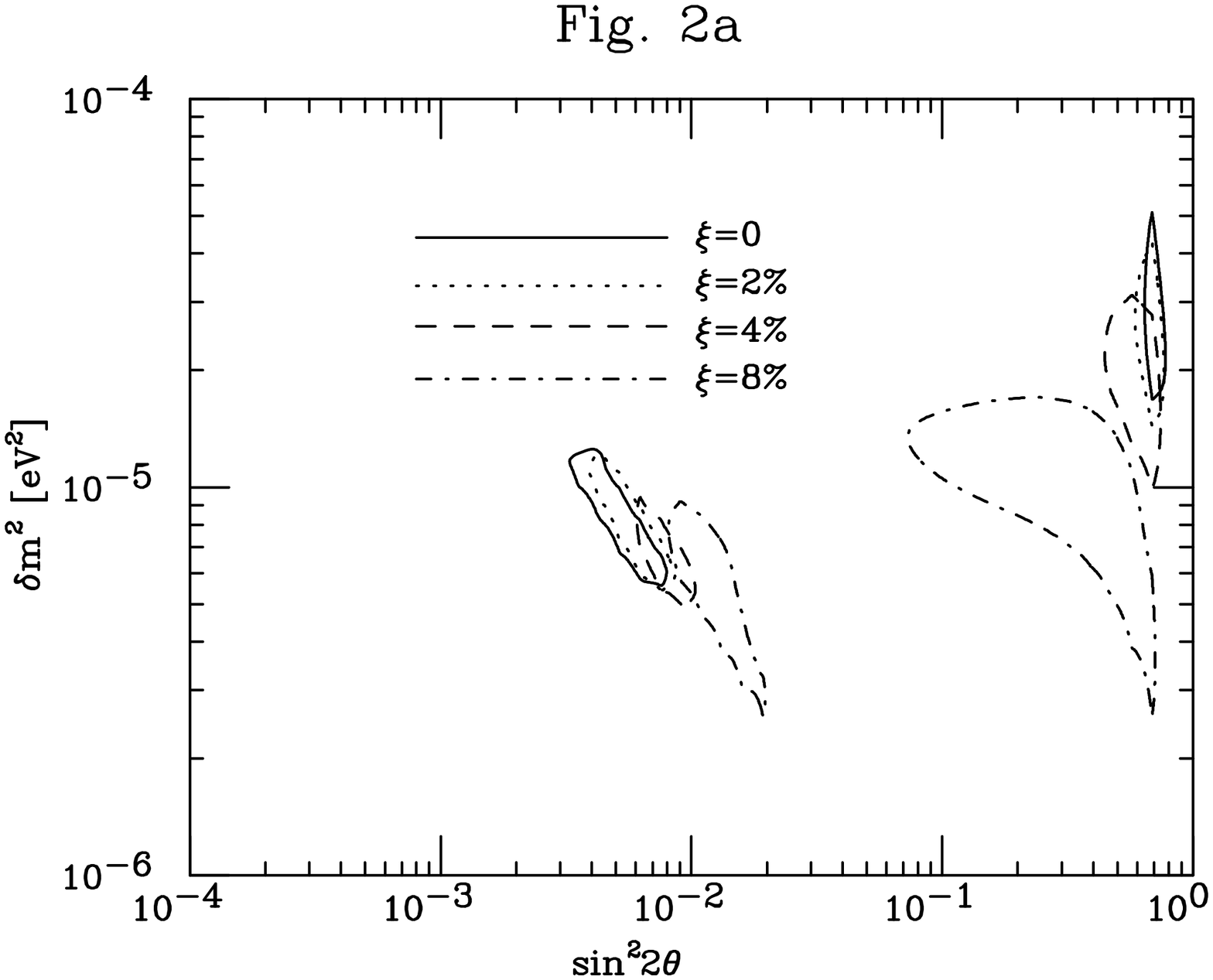,height=8.5cm,width=9.5cm,angle=90}
\vglue -8.5cm
\hglue 7cm
\psfig{file=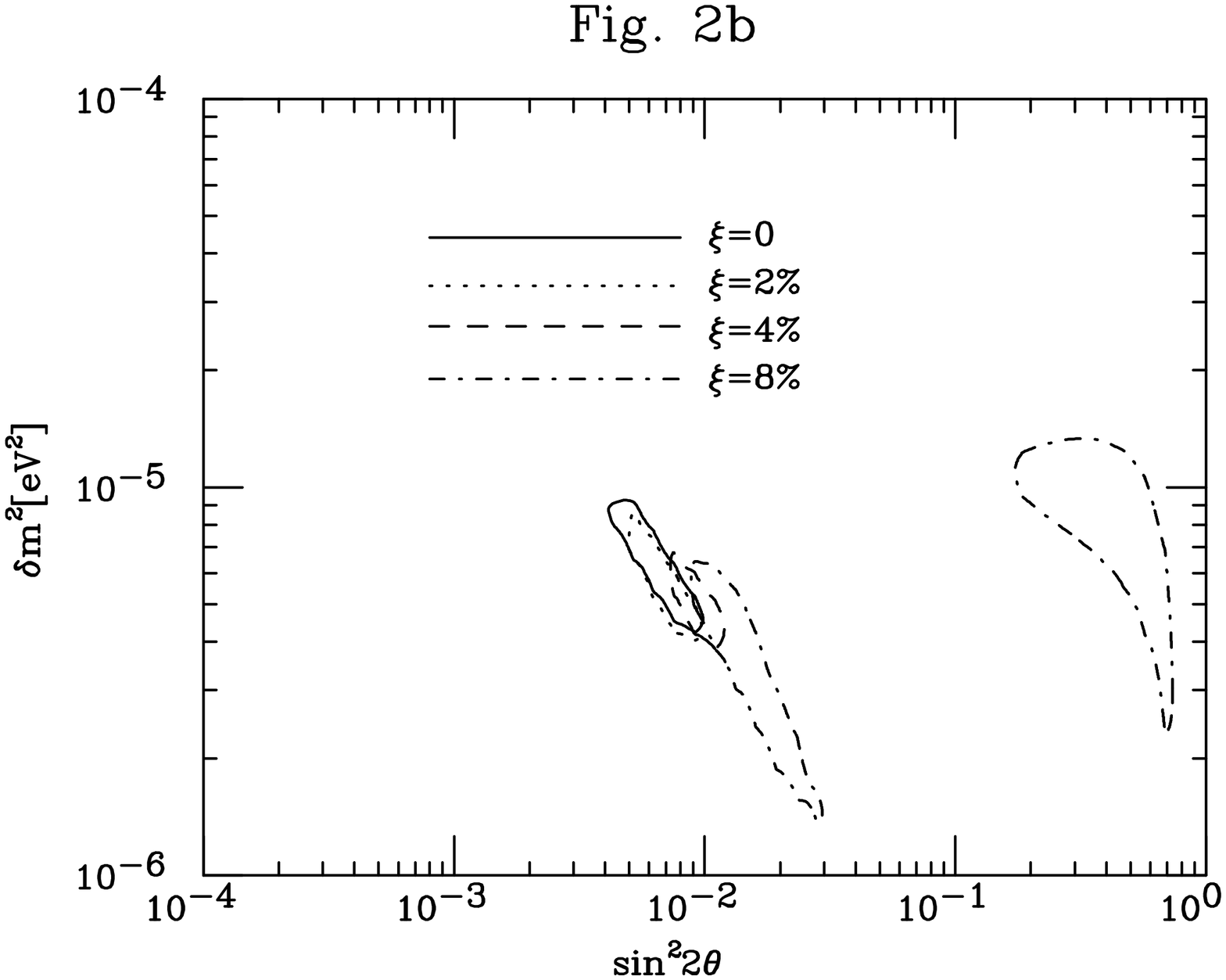,height=8.5cm,width=9.5cm,angle=90} 
\noindent
Fig. 2: The 90\% C.L. allowed regions for the $\nu_e\ra
\nu_{\mu,\tau}$ (Fig. 2a) and for the $\nu_e\ra
\nu_{s}$ (Fig. 2b) conversion. The different  curves refer to 
different values of matter noise level $\xi$ as indicated.
\newpage
%%%%%%%%%%%%%%%%%%% Fig. 3a - 3d %%%%%%%%%%%%%%%%%%%%%%%%%%%%%%%
\hglue -1.5cm
\psfig{file=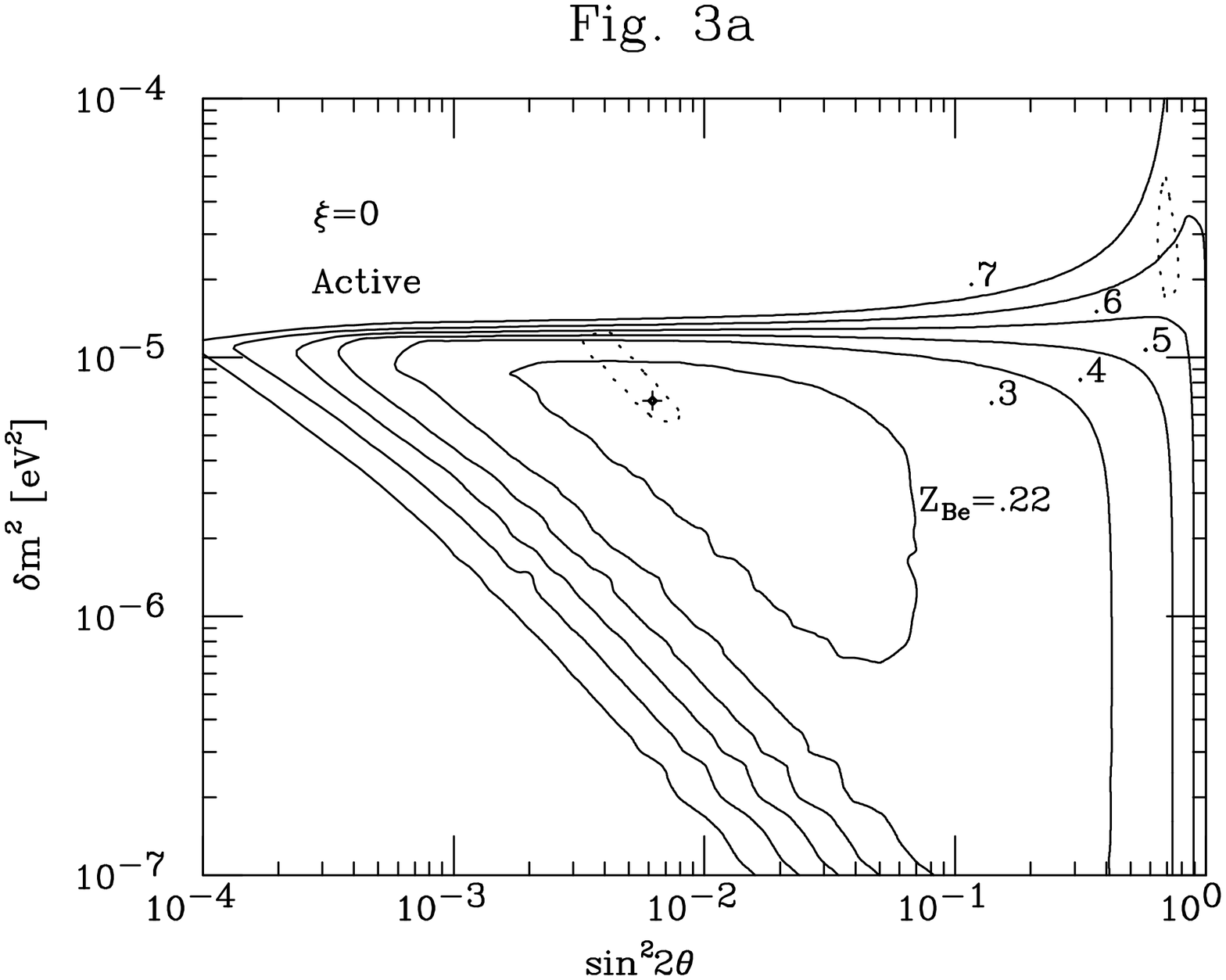,height=8.5cm,width=9.5cm,angle=90}
\vglue -8.5cm 
\hglue 7cm
\psfig{file=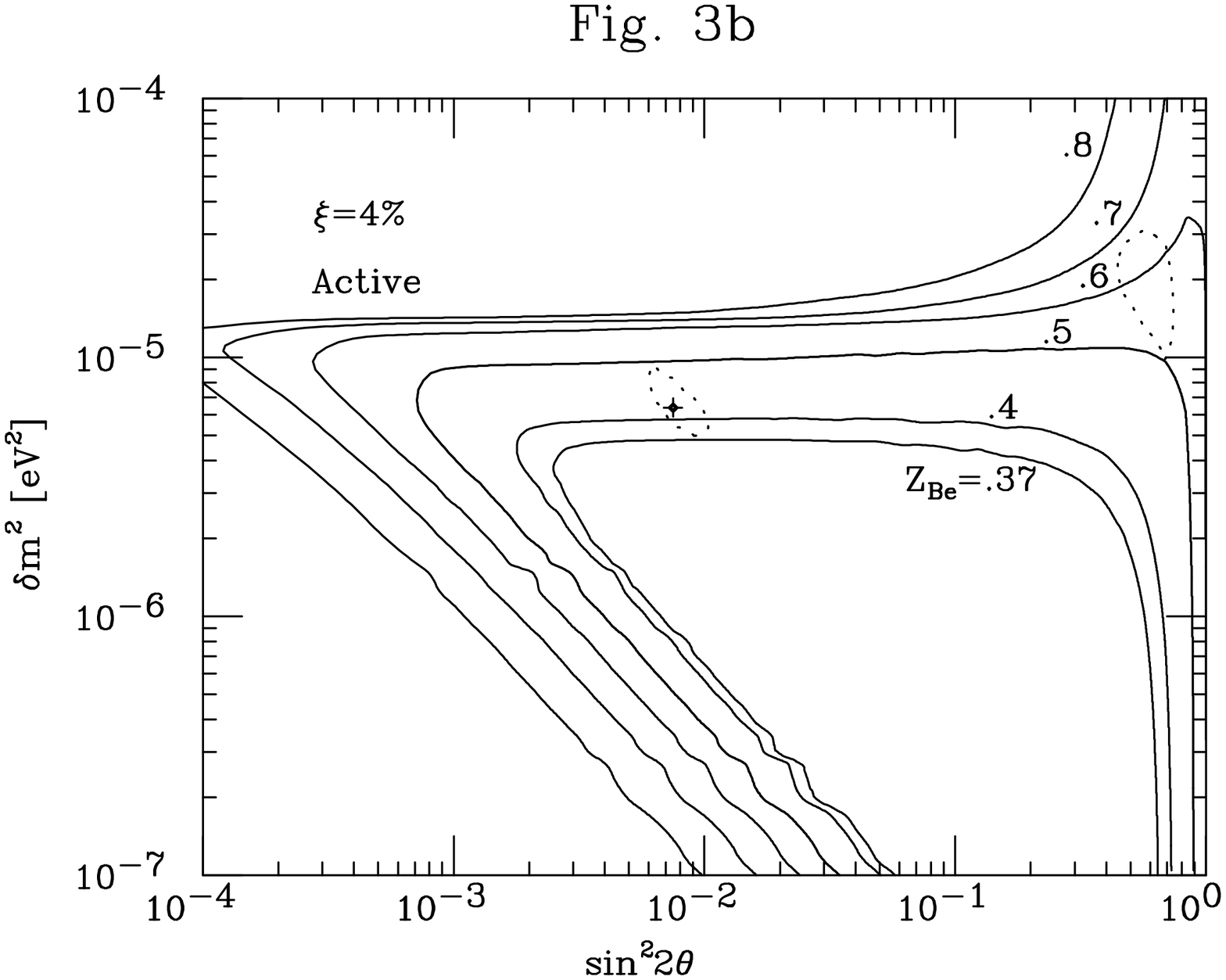,height=8.5cm,width=9.5cm,angle=90} 
%%%%%%%%%%%%%%%%%%% Fig. 3a - 3d %%%%%%%%%%%%%%%%%%%%%%%%%%%%%%%
\vglue 0.2cm
\hglue -1.5cm
\psfig{file=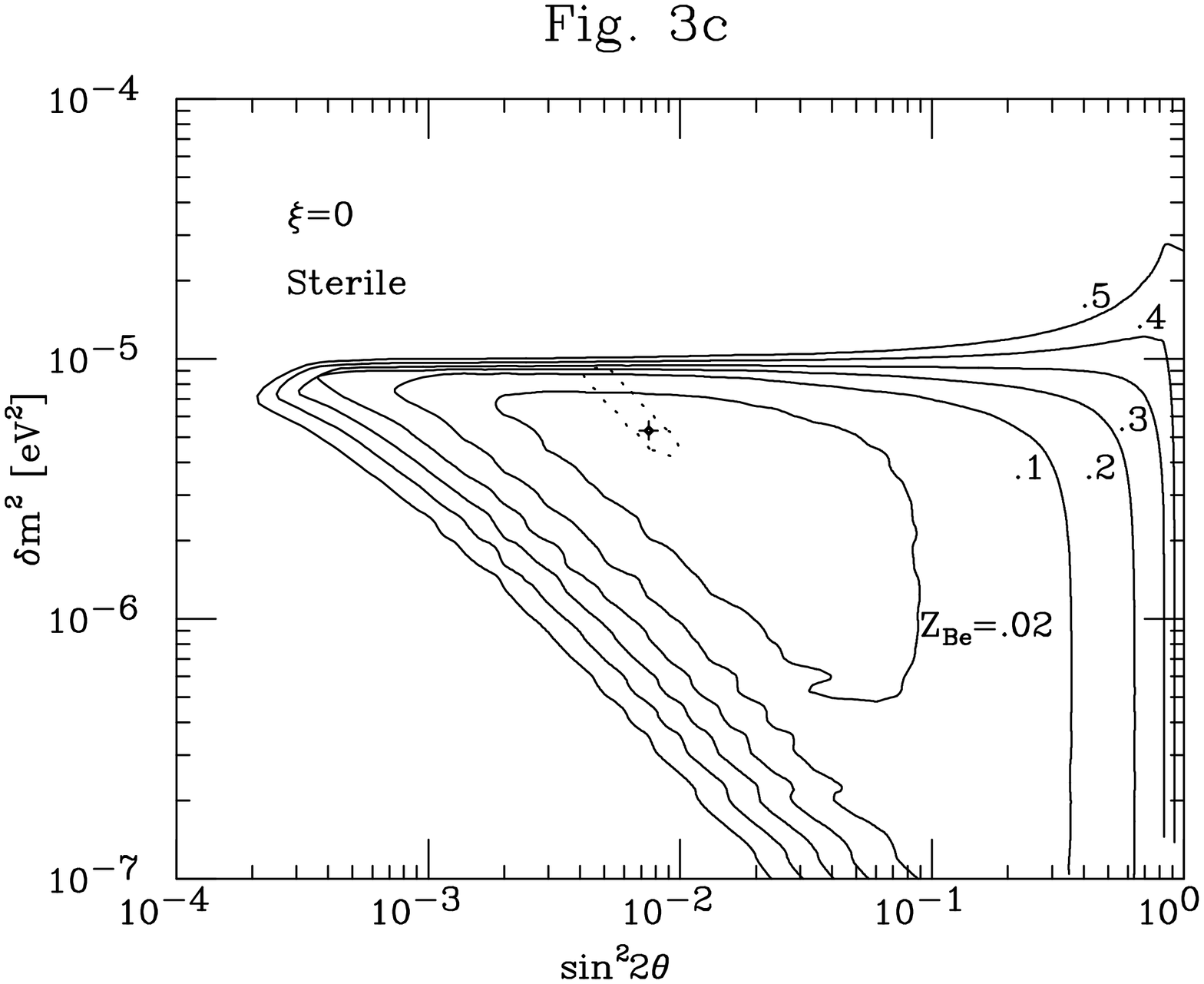,height=8.5cm,width=9.5cm,angle=90}
\vglue -8.5cm
\hglue 7cm
\psfig{file=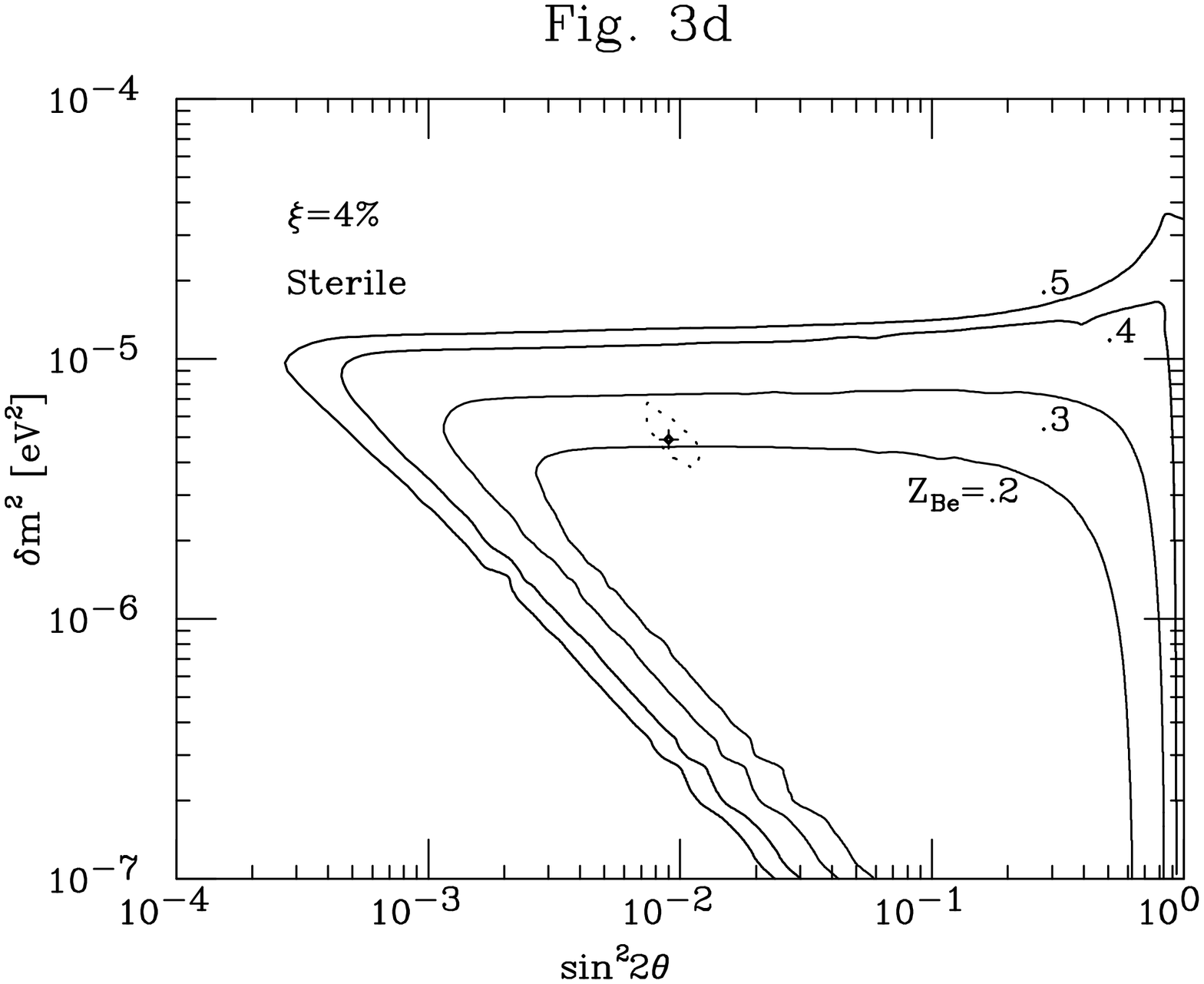,height=8.5cm,width=9.5cm,angle=90} 
\vglue 0.5cm
\noindent
Fig. 3: The iso $Z_{Be}= R^{pred}_{Be}/R^{BP95}_{Be}$ 
contours (figures at curve) in the $\nu-e$ scattering 
Borexino detector (solid lines). 
The threshold energy for the recoil electron detection is 0.25 MeV. 
The 90\% C.L. regions (dotted line) and the corresponding best fit point 
are also drawn. Fig. 3a and Fig. 3b refer to the case of 
$\nu_e\ra \nu_{\mu,\tau}$ conversion and for  $\xi=0$ and $\xi=4\%$, 
respectively.  Fig. 3c and Fig. 3d refer to the case of
$\nu_e\ra \nu_{s}$ conversion and for  $\xi=0$ and $\xi=4\%$, 
respectively.  
%%%%%%%%%%%%%%%%%%% Fig. 3a and 3b %%%%%%%%%%%%%%%%%%%%%%%%%%%%%%%

\vspace{0.5cm}

\begin{thebibliography}{99}

\bibitem{NRSV} 
H. Nunokawa, A. Rossi, V. Semikoz and J. W. F. Valle, preprint 
FTUV/95-47, IFIC/95-49, hep-ph/9602307, accepted for 
the publication in {\em Nucl. Phys.} {\bf B}.

\bibitem{cl}
B.T. Cleveland {\it et al.}, \nps{38}{95}{47}. 

\bibitem{ga}
GALLEX Collaboration,  P. Anselmann {\it et al.}, 
LNGS Report 95/37 (June 1995). 

\bibitem{sa}
SAGE  Collaboration,  J.S. Nico {\it et al.}, 
{\em Proc. 27th Conf. on High Energy Physics}, Glasgow, UK (July 1994). 

\bibitem{k}
Y. Suzuki, \nps{B38}{95}{54}

\bibitem{SSM}
J. N. Bahcall and R. K. Ulrich, \rmp{60}{90}{297}; \\
J. N. Bahcall and M. H. Pinsonneault, \rmp{64}{92}{885}; 
J. N. Bahcall and M. H. Pinsonneault, preprint IASSNS-AST 95/24

\bibitem{CF}
V. Castellani, {\it et al} \pl{B324}{94}{245};\\
N. Hata, S. Bludman, and P. Langacker, \pr{D49}{94}{3622};\\
V. Berezinsky, {\rm Comments on Nuclear and Particle Physics} {\bf 21} 
(1994) 249; \\
J. N. Bahcall, \pl{B338}{94}{276}. 

\bibitem{BFL}
V. Berezinsky, G. Fiorentini and M. Lissia, 
\pl{B365}{96}{185}.
 
\bibitem{MSW}
S. P. Mikheyev and  A. Yu. Smirnov, \sjnp{42}{86}{913};
{\em Sov. Phys. Usp.} {\bf 30} (1987) 759; \\
L. Wolfenstein, \pr {D17}{78}{2369};\ib{D20}{79}{2634}.

\bibitem{FIT}
G. Fiorentini {\it et al.} {\em Phys. Rev.} {\bf D49} (1994) 6298;\\
N. Hata and P. Langacker, {\em Phys. Rev.} {\bf D50} (1994) 632.

\bibitem{smirnov}
P. I. Krastev and A. Yu. Smirnov, {\em Phys. Lett.} {\bf B338} 
(1994) 282; \\ 
V. Berezinsky, G. Fiorentini and M. Lissia, 
{\em Phys. Lett.} {\bf B341} (1994) 38. 

\bibitem{Cala}
E. Calabresu {\it et al.}, \pr {D53}{96}{4211};\\
J. N. Bahcall and P. I. Krastev, Princeton preprint IASSSNS-AST 95/56,
hep-ph/9512378
 
\bibitem{KS}
P. I. Krastev and A. Yu Smirnov, {\em Phys. Lett.} {\bf B226} (1989) 341; 
{\em Mod. Phys. Lett.} {\bf A6} (1991) 1001. 

\bibitem{AbadaPetcov}
A. Schafer and S. E. Koonin, {\em Phys. Lett.} {\bf B185} (1987) 417; \\
R. F. Sawyer, {\em Phys. Rev.} {\bf D42} (1990) 3908; \\
A. Abada and S.T. Petcov, {\em Phys. Lett.} {\bf B279} (1992) 153.

\bibitem{BalantekinLoreti}
F. N. Loreti and A. B. Balantekin, \pr{D50}{94}{4762}.

\bibitem{dal}
J. Christensen-Dalsgaard, private communication.


\bibitem{gmode}
P. Kumar, E. Quataert, and J. N. Bahcall, astro-ph/9512091


\bibitem{turck}
S. Turck-Chi$\acute{e}$ze and I. Lopes, {\em Ap. J.} {\bf 408} (1993) 346; \\ 
S. Turck-Chi$\acute{e}$ze {\em et al.}, {\em Phys. Rep.} {\bf 230} (1993) 57. 

\bibitem{CDF}
V. Castellani, S. Degl'Innocenti and G. Fiorentini, 
{\em Astron. Astrophys.} {\bf 271} (1993) 601.

\bibitem{borex} 
C. Arpesella {\it et al.} (Borexino Collaboration), 
Proposal of BOREXINO (1991).

\end{thebibliography}
\end{document}